\documentclass[twocolumn]{revtex4}
\usepackage{latexsym}
\usepackage{amsmath}
\usepackage{times}
\usepackage{amssymb}
\usepackage{fancyheadings}
\usepackage[T1]{fontenc}
\usepackage[utf8]{inputenc}
\usepackage{url}
\usepackage{hyperref}
\usepackage{color}
\usepackage{graphicx}
 \usepackage{epsfig}

\begin{document}

\title{\bf The influence of architecture on collective charge transport
  in nanoparticle assemblies revealed by the fractal time series and
  topology of phase space manifolds}
\author{Bosiljka Tadi\'c$^{1}$}\thanks{Corresponding author. Email: bosiljka.tadic@ijs.si,  Tel.: +38614773767.}
\author{Miroslav Andjelkovi\'c$^{1,2}$}
\author{Milovan \v{S}uvakov$^3$ \vspace*{3mm}}
\affiliation{$^1$Department of Theoretical Physics, Jo\v zef Stefan Institute,
Jamova 39, Ljubljana, Slovenia }
\affiliation{$^2${Institute of Nuclear Sciences Vin\v ca,  University
    of Belgrade,  11000 Belgrade, Serbia}}
\affiliation{$^3$Institute of Physics,  University of Belgrade, Pregrevica 118, 11080
  Zemun-Belgrade, Serbia\vspace*{3mm}}


\begin{abstract}
\noindent 
Charge transport within Coulomb blockade regime in two-dimensional nanoparticle arrays exhibits nonlinear I-V characteristics, where the level of nonlinearity strongly associates with the array's architecture. Here, we use different mathematical approaches to quantify collective behavior in the charge transport inside the sample and its relationship to the structural characteristics of the assembly and the presence of charge disorder.
In particular, we simulate single-electron tunneling conduction in several assemblies with controlled variation of the structural components (branching, extended linear segments) that influence the local communication among the conducting paths between the electrodes.
Furthermore, by applying the fractal analysis of time series of the number of tunnelings and the technique of algebraic topology,  we unravel the temporal correlations and structure of the phase-space manifolds corresponding to the cooperative fluctuations of charge.
By tracking the I-V curves in different assemblies together with the indicators of collective dynamics and topology of manifolds in the state space, we show that the increased I-V nonlinearity is fully consistent with the enhanced aggregate fluctuations and topological complexity of the participating states.
The architecture that combines local branching and global topological disorder enables the creation of large drainage basins of nano-rivers leading to stronger cooperation effects.  Also, by determining shifts in the topology and collective transport features, we explore the impact of the size of electrodes and local charge disorder. The results are relevant for designing the nanoparticle devices with improved conduction; they also highlight the significance of topological descriptions for a broader understanding of the nature of fluctuations at the nanoscale.

\vspace*{2ex}\noindent\textit{\bf Keywords}: Complex Systems,
Nanoparticle assemblies, Single-electron tunneling, Collective charge
fluctuations, Nanonetworks, Numerical Methods, Fractal analysis, Algebraic topology.
\\[3pt]
\\[3pt]
\end{abstract}

\maketitle

\thispagestyle{fancy}

\section{Introduction\label{sec-intro}}
In the science of complex systems, understanding the emergence of distinct properties on a larger scale is one of the  central problems, which requires the use of advanced mathematical and numerical approaches
\cite{advances-book2011}. Nanostructured materials are examples of complex systems exhibiting new functionality at the assembly level \cite{philip-review,CB-mollinkers-emergence-2015}. The concepts of nanonetworks (\cite{we-nanonets} and references there) facilitate the application of graph theory methods for a quantitative study of complexity at a nanoscale. Here, we combine graph theory techniques with additional mathematical methods to investigate the connection between the cooperative electron transport through nanoparticle assemblies at applied bias within the Coulomb blockade regime and the architecture of the assembly.

In conducting nanoparticle arrangements, the Coulomb blockade conditions provide the  single-electron tunneling (SET) conduction  between neighbouring nanoparticles (\cite{book-tn-97,CB-granularmetals-revie2007,we-review} and references within).  The prototypes of systems with the Coulomb blockade transport are self-assembled 3-dimensional arrays \cite{CB-3DarrayNP-science2004} and nanoparticle films on substrates \cite{we-NL2007,we-Oxford}, consisting of small metallic nanoparticles with capacitative coupling along the tunneling junctions.
Recently,  similar conduction mechanisms have been described in quantum dot arrays in the reduced graphene oxide \cite{khondaker-RPB2011,khondaker-PhysChemC2013}, a new electronic and optoelectronic material \cite{RGO-review}.  In this case, quantum dots of graphene are separated by non-conducting areas through which the electrons can tunnel at the applied voltage bias.
The relevance of the Coulomb blockade transport has been experimentally  investigated in a variety of other nanostructures including nanowires \cite{current-NW}, granular metals
\cite{CB-1Dwd-granularmetalsTh2005}, and  thick films \cite{thick-films} as well as different molecular arrays \cite{CB-molecular-film-Nat2000,CB-moleculararray2010}.
At low temperatures, SET  represents a main process in the assemblies of small nanoparticles arranged with a fixed pattern of tunneling junctions \cite{mw-ctasmd-93,rhc-sbcvcotdasmd-95,wcycbhk-cbdtpgcs-97,plj-etmna-01,plerj-ptnrt-04,we-NL2007,we-Oxford} and graphene quantum dots  \cite{khondaker-RPB2011,khondaker-PhysChemC2013}.
Beside tunneling, another dynamical regimes, occurring in hybrid nanocomposites \cite{semi-metal2013} and separate time scales due to the motion of molecular linkers \cite{np-molecular2013},  have been investigated.

In the nanoparticle films, at low temperatures and the applied weak
bias at the electrodes encasing the array of nanoparticles
the current--voltage characteristic is nonlinear in a range of
voltages $V$ above a threshold $V_T$,
\begin{equation}
I(V)\sim (V-V_T)^\zeta  \ .
\label{eq-IV}
\end{equation}
It has been recognized that the degree of nonlinearity, which is
measured by the exponent $\zeta\in (1,5)$,  robustly correlates with
the structure of nanoparticle films  \cite{we-NL2007,we-Oxford,khondaker-PhysChemC2013} and their thickness \cite{thick-films,CB-coopertunnelingExp-2008}.
The origin of this phenomenon is in the \textit{cooperative charge transport} that involves multi-electron processes along the conduction paths, dynamically emerging between the electrodes.
Precisely, the SET conduction through a single Coulomb island between
the electrodes results in a linear $I(V)$ dependence. Similarly, $\zeta \simeq 1$ is found in one-dimensional
chains of nanoparticles supporting  a single conduction path
\cite{mw-ctasmd-93,we-review}. A detailed theoretical analysis of the tunneling
conduction through a chain of metallic grains with charge
disorder can be found in \cite{CB-1Dwd-granularmetalsTh2005}.
In contrast, multiple conduction paths establish between the electrodes in the case of
two-dimensional arrays and thick films.  Depending on the structural characteristics
of the array at local and global scale, such paths form drainage basins. Consequently, the cooperative
tunneling events may occur in such basins involving several conducting paths, which increase current through the system and leads to enhanced nonlinearity in $I(V)$ curves \cite{we-NL2007,CB-coopertunnelingExp-2008}.
While fluctuations of the current  are standardly measured at the electrode
\cite{mw-ctasmd-93,rhc-sbcvcotdasmd-95,plj-etmna-01,we-NL2007,khondaker-PhysChemC2013,thick-films,zabet-review2008,CB-coopertunnelingExp-2008},
a direct observation of the tunneling  events inside the sample remains a challenging problem to the experimental techniques
\cite{set-imaging}.  Therefore, the genesis of the collective charge fluctuations and
its connection with the structure of the array remains in the domain of numerical modeling.
In this regard, the idea of nanonetworks \cite{we-nanonets} provides the framework to quantify the structure of topologically disordered nanoparticle assemblies by graph theory methods \cite{we-iccs2006}. Furthermore, the numerical implementation of SET on an array of the \textit{arbitrary structure represented as a nanonetwork} has been introduced  \cite{MSBT_simulations2007,we-review}.

\begin{figure}[htb]
\begin{tabular}{cc} 
\resizebox{16pc}{!}{\includegraphics{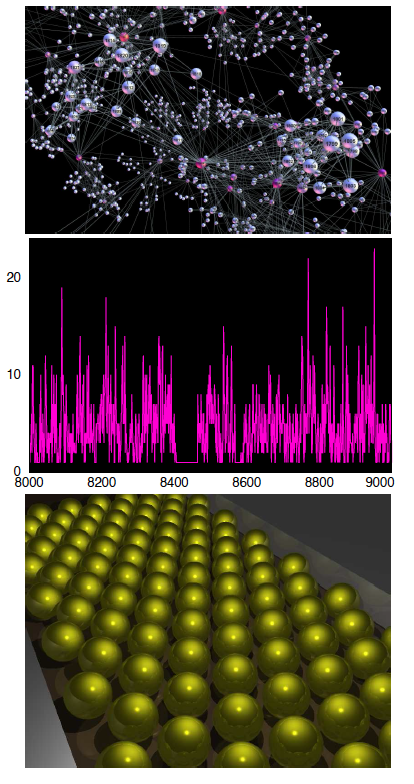}}\\
\end{tabular}
\caption{Illustration of three levels of the charge transport modeling.
  Bottom: Simulations of SET processes on nanoparticle array  between the electrodes; Middle:  Fractal analysis of the time series of the number
of tunnelings per time unit in the assembly; Top: Algebraic topology
analysis of the graph representing the phase-space manifold behind the
sequence of events in the time series.}
\label{fig-schematic}
\end{figure}

In this work, we combine different numerical techniques to study
collective features of charge transport through two-dimensional
nanoparticle assemblies; the aim is to examine cooperative dynamical
behavior involving different conduction paths in
connection with the structural elements of the assembly and the size
of electrodes as well as the effects of charge disorder.   Our
approach consists of three  levels illustrated in
Fig.\ \ref{fig-schematic}, relating to a suitable
mathematical modeling (see Methods for a detailed description). 
In particular:  (i) We construct several two-dimensional assemblies of nanoparticles connected by
tunneling junctions of a given structure. Then, setting the electrodes
(cf. an example in Fig.\ \ref{fig-schematic} bottom) and slowly
ramping the voltage bias, we simulate SET through such assemblies.  
(ii) We sample the time series of the relevant observables and perform
the fractal analysis of these time series, as described in Methods, to
determine the quantitative indicators of aggregate fluctuations.  Here, we consider temporal fluctuations of the number of tunnelings per time unit, $N_Q(t_k)$, an example is shown in Fig.\
\ref{fig-schematic} middle. 
(iii) Given recent developments of time-series--graphs duality
\cite{stoch-visibility,visibility,we-PRE}, we convert these time series into graphs.  The sets of data points specify a manifold
in the phase space of the system's states that are involved in the
course of events; the resulting graph then contains connections
between these states.  To explore the connection complexity among the
system's states, we use algebraic topology techniques (see Methods),
and study higher order combinatorial spaces (simplexes) of these graphs. 
Our comparative analysis of different nanoparticle assemblies reveals
a consistent correlation between the occurrence of collective charge
transport, topological complexity of the phase space manifolds, and the
$I(V)$ nonlinearity.

\section{Theoretical Background and $I(V)$ Nonlinearity of Different Assemblies\label{sec-theory}}
The system of small metallic nanoparticles arranged on a substrate
conducts when the applied bias exceeds a global threshold $V_T$, which is characteristic of the Coulomb blockade regime
\cite{book-tn-97,CB-coopertunnelingExp-2008,we-review}. 
In the Coulomb blockade conditions, the charging energy of a nanoparticle by a single electron $\Delta E_s=e^2/2C \gg k_BT$, where $C$ is the capacitance of the nanoparticle. The electrons, which are localised on nanoparticles, are transported by tunnelings through the junction between two nanoparticles when the voltage increases to exceed the Coulomb blockade voltage. The tunneling resistance satisfies the condition $R_t\gg h/e^2$  and the characteristic time scale is given by $\Delta t \sim R_tC$.   The array of nanoparticles thus represents a capacitative coupled system, as described in the original work \cite{mw-ctasmd-93}.  In \cite{we-review} this theoretical concept is generalised to consider an arbitrary structure of the nanoparticle system,  which is described by the adjacency matrix $A_{ij}$ of the underlying nanonetwork. In particular, the electrostatic energy $E$ of the assembly can be written in the matrix form as the follows
\begin{equation}
E=\frac{1}{2} \mathbf{Q}^\dag \mathbf{M}^{-1} \mathbf{Q} +  \mathbf{Q} \cdot
\mathbf{V}^{ext}  + Q_{\mu} \Phi^{\mu} \quad .
\label{eq-energy}
\end{equation}
where summation over $\mu$ applies and the potential corresponding to
the electrodes has the components
\begin{equation}
V^{ext}_i=\sum_{j,\mu}M_{ij}^{-1} C_{j,\mu} \Phi^{\mu} .
\label{eq-Vext}
\end{equation}
 Here, $\mathbf{Q}$ represents the vector of charges and $\mathbf{\Phi}$ the potential of nanoparticles
$i=1,2,\cdots N$. They satisfy the following relation at each nanoparticle:
\begin{equation}
Q_i=\sum_j C_{ij}(\Phi_i-\Phi_j) + \sum_\mu
C_{i,\mu}(\Phi_i-\Phi_\mu) 
\label{eq-Q-Phi}
\end{equation}
where $C_{ij}=CA_{ij}$ represents the capacitance between neighbouring
nanoparticles. The adjacency matrix element $A_{ij} =1$,
when the particles $i\leftrightarrow j$ are within the tunneling distance while $A_{ij} =0$,  when the distance between the particles is larger than the tunneling radius for the considered type of nanoparticles
\cite{we-review,zabet-review2008}. The index  $\mu \in \lbrace
+,-,g\rbrace$  stands for the positive and negative electrode
and the gate, respectively. 
Thus, the elements of the capacitance matrix $\mathbf{M}$ depend on the structure of the assembly: 
\begin{equation}
M_{ij}=\delta_{ij}\left(\sum_j C_{ij}+\sum_\mu C_{\mu,i}\right) -
C_{ij} .
\label{eq-M}
\end{equation}
Then  the potential at a nanoparticle  $i$ can be expressed as
\begin{equation}
\Phi_i= \sum_jM_{ij}^{-1}Q_j + V_i^{ext}
\label{eq-PhiQ}
\end{equation} 
in terms of all coupled charges and the potential at the electrodes.

The system is driven by slowly increasing the voltage bias between
the electrodes. Thus, the tunnelings start between the high-voltage electrode and the first layer of nanoparticles, cf.\ Fig.\ref{fig-schematic}, and gradually pushing through the sample the front reaches the other electrode when $V\simeq V_T$. Then the current through the sample can be measured. 
Due to the long-range electrostatic interactions, every tunneling event affects the potential--charge relation (\ref{eq-PhiQ}) away from the event junction. Consequently, the energy change in the whole assembly occurs.  Considering a particular tunneling event $a\to b$ from the nanoparticle $a$ to its neighbour nanoparticle $b$, the charge $Q_i$, measured in the number of electrons $e$, at a nanoparticle $i$  changes as
\begin{equation}
Q_i^\prime = Q_i + \delta_{ib} -\delta_{ia} \ , 
\label{eq-qprime}
\end{equation} 
causing the energy change of the assembly $\Delta E_{a\to b}
=E(\mathbf{Q^\prime}) - E(\mathbf{Q})$. As discussed in detail in
Ref.\ \cite{we-review}, introducing the following quantity
\begin{equation}
V_c= \sum_iQ_iM_{ic}^{-1} \ ,
\label{eq-Vc}
\end{equation}
allows to write the energy change of the assembly due to the tunneling
$a\to b$ in a concise form
\begin{eqnarray}
\Delta E_{a\to b} =\frac{1}{2}\left(M_{aa}^{-1} +M_{bb}^{-1}
  -M_{ab}^{-1} -M_{ba}^{-1}\right) \; \\
+ V_b -V_a + V_b^{ext}-V_a^{ext} ,
\label{eq-DE}
\end{eqnarray}
 and similarly  for tunnelings between the electrodes and a
 nanoparticle $a$
\begin{equation}
\Delta E_{a\leftrightarrow \pm} = \pm V_a + \frac{1}{2}M_{aa}^{-1} .
\label{eq-aEl}
\end{equation}

Theoretically, for a given value of the external voltage $V$, the
tuneling rate $\Gamma _{a\to b}$ is determined by the energy change
$\Delta E_{a\to b}$ according to the formula \cite{book-tn-97}
\begin{equation}
\Gamma _{a\to b} = \frac{1}{eR_{a\to b}}\frac{\Delta E_{a\to
    b}/e}{1-\mathrm{e}^{-\Delta E_{a\to b}/k_BT}} \ .
\label{eq-Gamma}
\end{equation} 
Consequently, the current through the junction between the
nanoparticles $a \leftrightarrow b$ is given by the balance of the
tunnelings
\begin{equation}
I_{a\leftrightarrow b}(V) = e\left[\Gamma_{a\to b}(V) -\Gamma_{b\to a}(V)\right] \ .
\label{eq-current}
\end{equation}
The total current across the sample is thus given by the 
tunneling rates  between the last layer of the nanoparticles and the
low-voltage electrode. Note that the voltage drop through
the sample according to (\ref{eq-PhiQ}) is nonlinear.

In the numerical implementation of the SET processes
\cite{we-review,MSBT_simulations2007}, the quantity $V_c$ defined by
(\ref{eq-Vc}) is computed recursively
\begin{equation}
a\to b : V_c^\prime = V_c + M_{bc}^{-1} -M_{ac}^{-1} \ , a\to \pm :
V_c^\prime = V_c \pm M_{ac}^{-1} .
\label{eq-recursive}
\end{equation}
 following every tuneling in the system. A more detailed description of the
 simulations is given in Methods.

\begin{figure}[htb]
\begin{tabular}{cc} 
\resizebox{14.8pc}{!}{\includegraphics{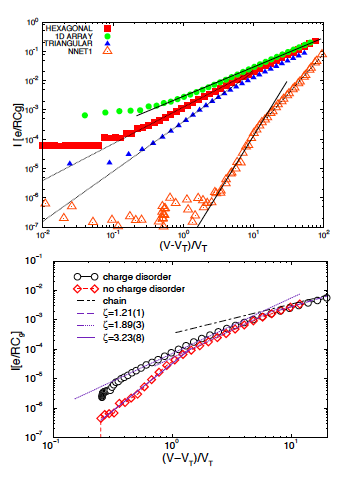}}\\
\end{tabular}
\caption{Current-Voltage curves for different nanoparticle
  assembles. Top: Curves from top to bottom correspond to
  one-dimensional chain, regular 
  hexagonal and  triangular arrays, and irregular structure NNET
  (cf. Fig.\ \ref{fig-nanonets});
  Bottom: The nanonetwork SF22 (cf. Fig.\ \ref{fig-nanonets}) in the presence of charge disorder (upper curve) and without charge disorder (lower curve).}
\label{fig-IV}
\end{figure}

As mentioned in the Introduction, the nonlinearity exponent in  I--V
characteristics, Eq.\ (\ref{eq-IV}),  strongly depends on the structure of the nanoparticle assembly
\cite{mw-ctasmd-93,rhc-sbcvcotdasmd-95,plj-etmna-01,we-NL2007,we-Oxford,khondaker-PhysChemC2013}. 
Moreover, it was observed that in the same structure 
the exponent $\zeta$ changes with variations of the distance between
the electrodes and their size \cite{we-Oxford,CB-electrodes}.
Our simulations of SET in different nanoparticle assemblies
\cite{we-NL2007,MSBT_simulations2007,we-review} confirm these experimental findings.  Fig.\ \ref{fig-IV} shows $I(V)$ curves obtained in different nanonetworks
including the structures studied in this work, cf. Fig.\ \ref{fig-nanonets}. 
For a single chain of nanoparticles,  the current--voltage
dependence is linear, i.e.,  $\zeta \simeq 1$, whereas for two-dimensional nanonetworks with a controlled structure  $\zeta
>2$, depending on the structure and the electrodes size.
Specifically, within numerical error bars, we find $\zeta=2.30$ and
$\zeta= 3.0$   in a regular triangular and hexagonal arrays,
respectively, with extended electrodes \cite{we-review}, in a good agreement with the experimental
results \cite{mw-ctasmd-93,plj-etmna-01}. An enhanced nonlinearity
 $\zeta =3.9$ characterises $I(V)$ curves in the highly irregular structure  possessing a combination of compact areas and large
voids (NNET in Fig.\ \ref{fig-nanonets}). The results agree well with the experimental studies of similar structures in \cite{we-NL2007}.  The curve is also shown in the top panel of Fig.\ \ref{fig-IV}.
The lower panel in Fig.\ \ref{fig-IV} shows $I(V)$ dependence in the
nanonetwork SF22. Its structure, also shown in Fig.\
\ref{fig-nanonets}, has a scale-free distribution of cells and a
fixed branching. The simulations of SET in this nanonetwork yield $\zeta =3.23$ for a narrow voltage range. However, we obtain a reduced exponent  $\zeta =1.89$ but an extended scaling region in the presence
of charge disorder. In general, charge disorder originates from fractional charges on the substrate \cite{mw-ctasmd-93,plj-etmna-01,CB-1Dwd-granularmetalsTh2005}), which  causes a weaker nonlinearity.
In the structure with nearly hexagonal cells, CNET in Fig.\
\ref{fig-nanonets}, the simulations performed in \cite{we-review} using varied sizes of the electrodes resulted in a weakly reduced exponent $\zeta \in [2.3,2.5]$. A  more recent study of regular arrays \cite{CB-electrodes} suggests that $\zeta$ decreases as a logarithm of the ratio between the size of electrode and their distance.

\section{Methods\label{sec-methods}}
\begin{figure*}[!htb]
\begin{tabular}{cc} 
\resizebox{32pc}{!}{\includegraphics{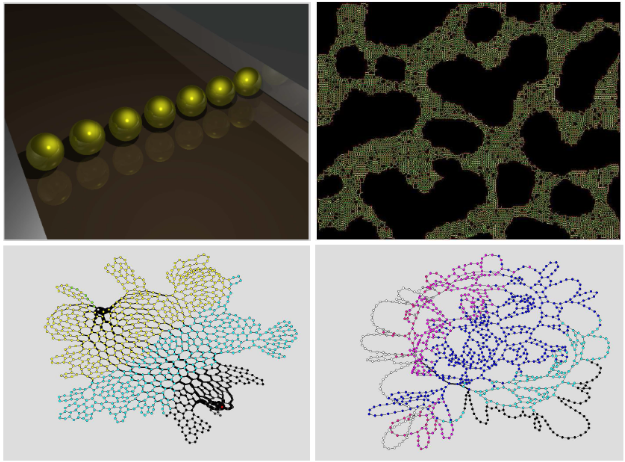}}&
\end{tabular}
\caption{Top left: a 3-dimensional rendering of a chain of nanoparticles on a substrate and the attached electrodes. Top right: A
2-dimensional projection of the nanonetwork NNET with 5000 nanoparticles self-assembled by the evaporation methods (data are from
Ref.\ \cite{we-NL2007}). The lines indicate the tunneling junctions while a nanoparticle is found at the intersections. Bottom:  CNET with
a point-size electrodes 
(left) and SF22 with the extended electrodes (right), both grown by
the cell aggregation 
\cite{we-iccs2006}.  The voltage drop across the sample is indicated
by the color of the  network's areas with high, intermediate and low potential.
}
\label{fig-nanonets}
\end{figure*}

The modeling approach on three levels mentioned in
the Introduction encompasses different scales and algorithms, from the
single-electron tunnelings at each junction of the nanonetwork in real
space to collective charge fluctuations of the whole assembly, which
are studied in time and the abstract phase space manifolds. 
The computational hierarchy is described as follows. \\
\textit{I. Simulations of SET on nanonetworks\label{sec-SET}.}
The SET processes are simulated in four different nanoparticles
assemblies represented by the nanonetworks in Fig.\ \ref{fig-nanonets}. First, the structure of the nanonetwork is included by specified adjacency matrix $\mathbf{A}$ and
the electrodes are set to some periphery nodes. Usually, we chose
$P/4$ nodes in the case of extended electrodes, or a single node, in
the case of point-size electrodes, at two opposite sides of the
network; here, $P$ is the number of periphery nodes of a given
structure (see Fig.\ \ref{fig-schematic}). 
Then the capacitance matrix and its inverse are computed and the vectors $\mathbf{Q}$ and $\mathbf{V_+}$ are initialized as zeros. 
To start the process, the time $t$ is initialised and the bias voltage set. For each value of the bias in the range
$V\in[0,V_{max}]$  the number of steps are
  performed as follows. At each step, a tuneling is attempted along
  each junction $i\to j$ and the values  $V_c$ for all nodes updated
  according to  (\ref{eq-recursive}) and the corresponding energy charges
$\Delta E_{i\to j}(t)$ are computed. Then, each tuneling
rate $\Gamma _{i\to j}(t)$ is determined according to
(\ref{eq-Gamma}); the delay time $\Delta t_{ij}$ of the tuneling along
the $i\to j$-junction is estimated.  Following the
description in \cite{we-review},  $\Delta t_{ij}= \frac{-log (1-x_{ij})
-\sum_{k>k_0}\delta t_k\Gamma_{i\to j}(t_k)}{\Gamma _{i\to
  j}(t)}$, where $t_{k_0}$ is the time of the last tunneling at the
junction and $x_{ij}$ is a uniform random number.
The tunneling along the junction with a minimal delay
time is processed and the time increased accordingly $t\to t+\Delta
t_{ij}$. Subsequently, the charge  $Q_i$ and $V_i$ are updated
for all nodes, and the process is repeated to find next tunneling event
and so on. 
The simulated data are in the limit $C/C_g=10^{-4}$, where a faster
algorithm for computing the inverse of the capacitance matrix can be
used \cite{we-review}. We also keep $T=0$ and $\Phi_-=\Phi_g=0$ while
$\Phi_+=V$, the applied voltage. The simulations are performed until
$V\gtrsim 10V_T$ in the corresponding nanonetwork.
To sample time series of interest for this work, we set an
appropriate time unit according to the average tunneling rate
$\delta t =a/\langle \Gamma _{ij}\rangle$ in each nanonetwork. The parameter $a=1$
in the one-dimensional array and takes different values\cite{we-review}
$a=$4.11 and  17.1 in CNET and NNET, respectively, in connection with the estimated number of potential paths in these irregular arrays. Then
the time series represents the sequence of the number of tunnelings
$N_Q(t_k)$,  $k=1,2 \cdots $ in the entire array per the identified time unit.

\textit{II. Fractal time series analysis and temporal correlations.\label{sec-TS}}
The occurrence of collective charge fluctuations manifests in
long-range temporal correlations, clustering of events and fractal features of time series. Various
indicators of the collective behavior are determined by analysis of the time series of the number of tunnelings $N_Q(t_k)$.  In
particular, the temporal correlations occurring in the streams of
events result in the  power spectrum  
\begin{equation}
S(\nu ) \sim \nu ^{-\phi} 
\label{eq-PS}
\end{equation}
with a power-law decay in a range of frequencies $\nu$. 
The avalanches of tunnelings accompany such temporal correlations. Using the standard numerical procedure 
\cite{tadic1999,we-SciRep2015}, the avalanches are identified in the stationary signal above $V_T$.
The size of an avalanche comprises the area below a regular part of the signal above the zero thresholds.  A broad distribution of avalanche sizes with a power-law tail suggests that, in a state with long-range correlations, the size of a triggered avalanche is not necessarily proportional to the triggering action. Moreover, the system's relaxation in the response to the external driving is characterized by the differences between the size of consecutive events ( first return). 
 The following expression, characteristic of the nonextensive statistical mechanics  \cite{tsallis-2009_intro,temporal-corr-AR2013,pavlos_q-returns2014},
\begin{equation}
P_\kappa(X)=A\left[1-(1-q_\kappa)\left(\frac{X}{X_0}\right)^\kappa\right]^{-1/1-q_\kappa} 
 \label{eq-q-exp}
\end{equation}
satisfactorily reproduces the distributions of the avalanche sizes,
i.e., $q_1$-exponential, for $\kappa =1$,  and  the returns
$q_2$-Gaussian, for $\kappa=2$.

According to the fractal analysis of complex signals \cite{MFRA-uspekhi2007}, the
time series profile $Y(i)=\sum_{k=1}^i(N_Q(t_k)-\langle
N_Q(t_k)\rangle$ is divided into $N_n$ segments of length $n$. The
standard deviation $F_2(\mu,n) =
\frac{\sum_{i=1}^n[Y((\mu-1)n+i)-y_\mu(i)]^2}{n}$ around the local
trend $y_\mu(i) $  is computed at each segment $\mu =1,2,\cdots N_n$.  
Then the average over all segments exhibits a scaling law
\begin{equation}
F_2(n)=(1/N_n)\sum_{\mu=1}^{N_n}F_2(\mu,n) \sim n^H 
\label{eq-F2}
\end{equation}
with respect to the varied segment length $n$, where $H$ is Hurst exponent. While $H=1/2$ characterises random
fluctuations, the values $H\in(1/2,1)$ indicate the persisten fluctuations of the fractional Gaussian noise signal \cite{fractalSPP,MFRA-uspekhi2007,dmfra-sunspot2006}.

\begin{figure}[!htb]
\begin{tabular}{cc} 
\resizebox{14pc}{!}{\includegraphics{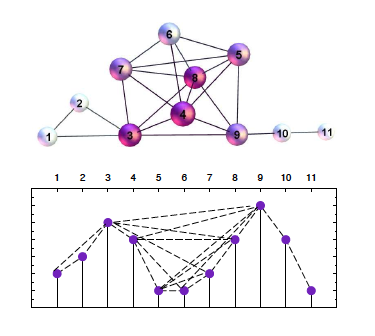}}\\
\end{tabular}
\caption{The sequence of data points in the lower pannel is
  mapped onto a graph in the upper pannel; each data point becomes a node of
  the graph, while the graph's edges are inserted according to the 
  'natural visibility' between data points, indicated by broken lines.}
\label{fig-visibility}
\end{figure}

\textit{III. Mapping time series onto QTS-graphs.\label{sec-visibility}}
The sequence of events captured by the time series of the number of
tunnelings represents a manifold in the state space of the underlying
nanonetwork. Dealing with a fractal signal (see Results), we expect a
more complex connections among these states to exist beyond the
actually realised sequence. To reveal such complexity of the system's phase space, we use the mapping of time series to mathematical
graphs recently developed \cite{visibility,stoch-visibility}. Here, we apply the 'natural visibility' mapping \cite{visibility,we-PRE}, which is particularly suitable in the case of persistent fluctuations \cite{we-PRE}. 
The mapping procedure is illustrated in Fig.\ \ref{fig-visibility}. Each data point of the time series is represented by a node of the graph, here
termed QTS-graph to indicate the charge fluctuations time series. The node is connected by undirected links with all other data points that are 'visible' from that data point, where the vertical bars are considered as non-transparent. Note that by varying the mapping procedure different graphs can be obtained. However, here we use the same mapping procedure for different time series (i.e., charge fluctuations in different nanonetworks) and perform a \textit{comparative analysis of the structure of the resulting graphs}.  For this purpose, equal parts
of the time series are mapped in each of the considered nanoparticle
systems, i.e.,   2000 data points,  skipping the initial 8000 points.

\textit{IV. Algebraic topology measures of  QTS-graphs.\label{sec-AT}}
Beyond standard graph theory measures \cite{bb-book,sergey-lectures}, the algebraic topology of graphs \cite{jj-book} is utilised to determine the higher-order structures of a graph, i.e., simplicial objects and their collections, simplicial complexes, closed under the inclusion of faces. Using the paradigm of Taylor expansion, such combinatorial spaces correspond to higher-order terms while the graph itself represents the linear term. 
Recently, the computational topology techniques have been applied  to
analyse the hierarchical organisation in online social networks
\cite{we-PhysA2015}, study of grain connectedness in trapped granular
flow \cite{AT-granular}, and to reveal the changes in the topological structure of state space across the traffic jamming regime \cite{we-PRE}.

The method that we use 
\cite{Bron,cliquecomplexes} identifies simplexes as maximal complete subgraphs or cliques. Thus, $q=0$  represent an isolated node, $q=1$ two nodes
connected with a link, $q=2$ is a triangle, $q=3$ a tetrahedron and so
on until $\sigma_{q_{max}}$, which identifies the highest order clique present
in the graph. Faces of a simplex $\sigma _q$ are the subsets $\sigma
_r< \sigma_q$.  The method determines cliques of all orders and
identifies the nodes that belong to each clique. Utilising this rich
information, we characterise the topological complexity of the graph at the global graph's level, i.e., by defining various structure
vectors, as well as the level of each node
\cite{we-PhysA2015,we-PRE}. In particular, having identified the
graph's topology layers $q=0,1,2,\cdots q_{max}$, we describe the
number of cliques and how they are interconnected via
shared faces at each level from $q_{max}-1, ...,1$:
\begin{itemize}
\item Three structure vectors of graph have the components $G_q$,
  $n_q$ and ${\hat{G}_q}=1-G_q/n_q$, which determine, respectively,
  the number of connected components at the level $q$, the number of
  simplexes from the level $q$ upwards, and the degree of
  connectedness between the simplexes at $q$-level.

\item The node's structure vector is defined \cite{we-PhysA2015}  by 
  the components $G^i_q$, the number of simplexes of order $q$ to which the node $i$
  participates. Then   $dim(G_i)=\sum_qG^i_q$ is the node's topological dimension.

\item The topological entropy $S_G(q)$ and the ``response'' function $f_q$ are
  defined \cite{we-PRE} using the above quantities. Namely, 
  the probability that a particular node $i$ contributes to the
  occupation of the topological level $q$ is  $p^i_q=G^i_q/\sum_i
  G^i_q$. Then the graph's entropy is
\begin{equation} 
S_G(q) = -\frac{\sum_ip_q^i
    log_{10}p_q^i}{log_{10}\sum_i(1-\delta_{G_q^i,0})} \ ,
\label{eq-entropy}
\end{equation}
where the sum in the 
  denominator indicates the number of nodes with a nonzero entry
  at the considered topology level. The $q$-level component $f_q$ is defined as the 
  \textit{number of simplexes and shared faces at the topology}
    level $q$.
\end{itemize}
 Notice that that the topological ``response''  $f_q$
 is different from the component  $n_q$ of the above defined second
 structure vector.   The study in  \cite{we-PRE} have
shown that the function $f_q$ precisely 
 captures the topology shifts occurring  in the underlying time series
 due to changed driving conditions.

\section{Results and Discussion\label{sec-results}}

\subsection{Nanonetworks with different structural components and the
  origin of cooperative charge transport\label{sec-nanonets}}
For a comparative analysis of charge transport, the simulations of SET
are performed in four nanoparticle assemblies with different
structural characteristics.  Specifically, two self-assembled
nanoparticle arrays on a substrate, which are represented by the
nanonetworks in the top row of Fig.\ \ref{fig-nanonets},  and two
computer-generated structures, in the bottom row. For this work, these structures are identified as follows. The
one-dimensional chain of nanoparticles between the electrodes, where
we assume  the presence of charge disorder $x_i\in [0,1]$ as a uniform random number, is named 1Dwd.  Further, NNET is a  strongly inhomogeneous structure of nanoparticles, which is self-assembled on the substrate by the evaporation process \cite{we-NL2007}. The occurrence
of empty areas of different sizes as well as the regions where the
nanoparticles appear to be densely packed is the marked characteristics
of this assembly. CNET, shown in the lower left,  and SF22 in the
lower right panel, are grown by the cell aggregation process \cite{we-iccs2006}.
In these networks, we control the appearance of 
two relevant structural elements---branching and the extended linear
segments---that appear statistically in the above described self-assembled structures.
Namely, CNET has nearly regular hexagonal cells, resembling the
dominant shapes in the dense areas of the NNET, while SF22 exhibits voids of different sizes; the cell sizes are taken from a
power-law distribution with the exponent 2.2 \cite{we-SFloops2006}. 
Moreover, both CNET and SF22 have a fixed degree of internal
nodes $k=3$, which reduces the branching possibilities for the
tunnelings, in contrast with the NNET, where a broader degree
distribution of nodes is found \cite{we-NL2007,we-review}.
Furthermore, the electrodes are set to the left and right edge of
the NNET, thus touching the closest layer of the
nanoparticles. Similarly, the extended positive and negative electrode are set each along a quarter of the periphery nodes, indicated by white and black nodes in SF22 structure in 
Fig.\ \ref{fig-nanonets}. In contrast, the point-size electrodes are
attached to two periphery nodes in CNET and two
nanoparticles at the opposite ends of the 1Dwd chain. 

The situation with point-size electrodes in CNET in Fig.\
\ref{fig-nanonets} readily illustrates the importance of local structure for the collective charge transport of the assembly. Increasing voltage bias permits tunnelings between the electrode and the first layer, consisting of two connected nanoparticles.  Then further tunnelings can occur by breaking the Coulomb blockade along the junctions towards nearest neighbor nanoparticles. 
The preferred direction of tunnelings follows the potential drop, which
is indicated by the different color of nodes. Apart from the long-range
electrostatic interactions, the local balance between the charge and
potential  (\ref{eq-PhiQ}) at each nanoparticle depends on its neighbourhood.
At $V\sim V_T$ the number of charges in the system is large enough to
allow the first conduction channel to form along the shortest path
between the electrodes. 
For $V\gtrsim V_T$ the conditions for the appearance of another next-to-shortest path are met. Here, the branching
possibilities for the charge flow play an important role. Consequently, the process can involve several paths thus making
a river-like structure that drains at the last layer of
nanoparticles.  These paths often share some central nodes and
junctions. The most used junctions (indicated by thick lines of CNET in
Fig.\ \ref{fig-nanonets}) make the main conduction channels 
\cite{we-review}, whose geometry crucially depends on the local structure. 
Draining along the conduction paths can cause a cascade of tunnelings,
where each event satisfies the local Coulomb blockade threshold and
charge--potential balance.  Thus, in this range of voltages,  the SET represents a dynamical percolation problem
\cite{plerj-ptnrt-04} where several conduction paths contribute to the
measured current, resulting in the nonlinear $I(V)$ curves in Fig.\ \ref{fig-IV}.   
The quantitative analysis presented in the remaining part of this
section confirms this picture. On the other hand, for 
$V\gg V_T$, the local potential exceeds the Coulomb blockade potential
yielding  Ohmic conduction and  the crossover to a linear $I(V)$ dependence in Fig.\ \ref{fig-IV}.

\subsection{Temporal correlations and the evidence of collective
  charge fluctuations\label{sec-collective}}
\begin{figure}[htb]
\begin{tabular}{cc} 
\resizebox{18pc}{!}{\includegraphics{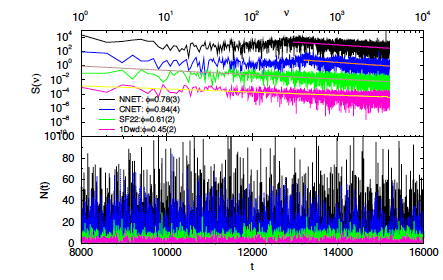}}\\
\end{tabular}
\caption{ Time series of the number of tunnelings (bottom panel) and
  the corresponding power spectrum (top panel) in four nanoparticle
  networks from Fig.\ \ref{fig-nanonets}. The legend and color apply
  to both panels. The signal for NNET is scaled by 1/5 to fit
the scale.}
\label{fig-TS-PS}
\end{figure}

The number of tunnelings per the identified unit interval is recorded from the SET simulations in four nanonetworks of 
Fig.\ \ref{fig-nanonets}. The resulting time series are shown in Fig.\ \ref{fig-TS-PS}.
The corresponding power spectra displayed in top panel of Fig.\
\ref{fig-TS-PS}, indicate the occurrence of temporal correlations that
depend on the structure of the underlying nanonetwork. Specifically,
the Eq.\ (\ref{eq-PS}) applies within a different range of frequencies and different exponent $\phi$, which is shown in the legend.  Apart from slight variations in the exponent, the power spectrum reveals the certain similarity between the charge fluctuations in the nanonetworks SF22 and 1Dwd, on one side, and NNET and CNET, on the
other. Hence, the dominance of the linear elements, which is obvious
to the chain structure, seems to play a role in the SF22 too. The
spectrum exhibits the power-law decay (\ref{eq-PS}) in the entire
range of $\nu$. On the other hand, plenty of branching possibilities
in CNET and NNET lead to the occurrence of multiple paths; a typical
scale appears, which leads to the peak in the spectrum between the
correlated high-frequency part and  the rest of the spectrum resembling a white noise. 

\begin{figure}[htb]
\begin{tabular}{cc} 
\resizebox{18pc}{!}{\includegraphics{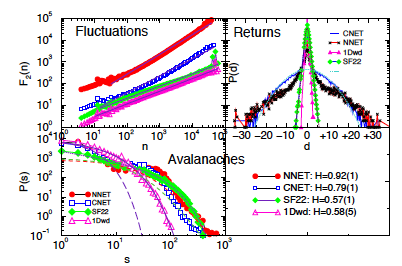}}\\
\end{tabular}
\caption{For four nanonetworks from Fig.\ \ref{fig-nanonets}:  Double
  logarithmic plot of the fluctuations $F_2(n)$ against the segment
  size $n$, top left, and the scaling regions indicated by a straight line on
  each curve;  The distributions $P(s)$ of the avalanche sizes $s$,
  bottom left, and the distribution $P(d)$ of the first returns  $d$
  of the signal.}
\label{fig-avalanches}
\end{figure}

A further similarity between these two groups of nanostructures is
found considering the nature of fluctuations and the statistics of avalanches,
which are displayed in Fig.\ \ref{fig-avalanches}. Computing the
standard deviations around a local trend on the segment of length $n$ of the time series, we find the scaling regions according to Eq.\
(\ref{eq-F2}) and determine the corresponding Hurst exponent. The values of the exponents for all time series,
which are listed in the legend, suggest that \textit{persistent}
fluctuations of charge transport occur in all considered
nanonetworks. The fluctuations in the linear chain of
nanoparticles and the SF22 structure have a similar value of the Hurst exponent, again suggesting the relevance of the
extended linear segments in these nanonetworks. Notably, the values of $H$ for these nanostructures are slightly larger than the case of the randomized time series, where $H=0.5$ within error bars. However, the two structures CNET and NNET, which allow the formation of more
communicating paths,  exhibit much stronger fluctuations
and consequently larger values of the Hurst exponent.  
Furthermore, the statistics of the returns $d\equiv  N_Q(t_k)-N_Q(t_{k-1})$ and the sizes  $s$ of clustered events
(avalanches) also suggests a similar grouping of these
nanostructures. On one side,  the charge fluctuations in SF22 and 1Dwd structures has the exponential distribution of avalanche sizes (with a larger cut-off in the case of SF22) and Gaussian
distributions of the returns.
On the other hand,  non-Gaussian fluctuations are found in the case
of CNET and NNET.  In this case, the distribution of avalanche sizes
exhibits a power-law tail compatible with the expressions
(\ref{eq-q-exp}) with $\kappa =1$ and
$q_1\sim 1.33$. 
Similarly,  the tails of the distribution of the returns can be fitted with
(\ref{eq-q-exp}), where $\kappa=2$, i.e., $q_2$-Gaussian, which is  often found in complex signals \cite{pavlos_q-returns2014}.
It should be stressed that the part of the data for small returns and small avalanches in these two nanonetworks virtually coincides with the ones found in the chain and SF22 structures. These weak fluctuations correspond to sporadic tunnelings away from the main conduction paths. This feature of the fluctuations is particularly pronounced in the case of CNET, where the most used area of the nanonetwork is reduced due to the point-size electrodes (cf.  Fig.\ \ref{fig-nanonets}).

\subsection{Topology of phase space manifolds related with the
  collective charge fluctuations\label{sec-topology}}
\begin{figure}[htb]
\begin{tabular}{cc} 
\resizebox{16pc}{!}{\includegraphics{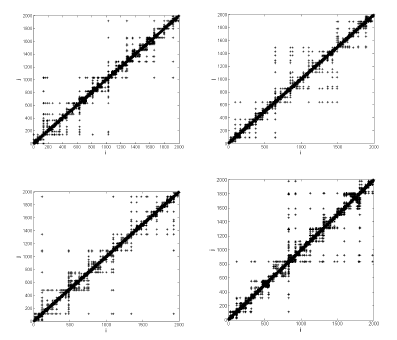}}\\
\end{tabular}
\caption{Adjacency matrix of the QTS-graph related with the charge
  fluctuations in the nanoparticle assemblies (a) 1Dwd,  (b) NNET, (c)
  CNET and (d) SF22.}
\label{fig-AMs}
\end{figure}

\begin{figure}[htb]
\begin{tabular}{cc} 
\resizebox{16.8pc}{!}{\includegraphics{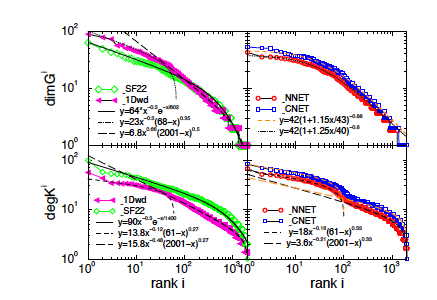}}\\
\end{tabular}
\caption{Ranking distributions of the node's degree (lower panels) and
  topological dimension (upper panels) in the studied QTS graphs
  of charge fluctuations in four nanonetworks from Fig.\
  \ref{fig-nanonets}. }
\label{fig-ranked-nodes}
\end{figure}

\begin{figure}[htb]
\begin{tabular}{cc} 
\resizebox{18pc}{!}{\includegraphics{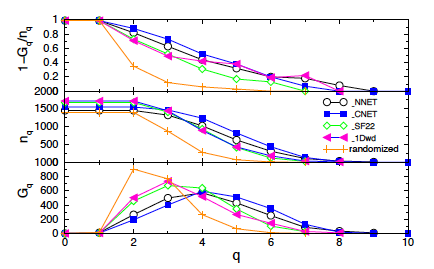}}\\
\end{tabular}
\caption{Components of the 1st, 2nd and 3rd structure vector, $G_q=1-G_q/n_q$,
  $n_q$ and ${\hat{G}_q}$, respectively, plotted against the topology
  level $q$ in the QTS-graphs related with the time series of charge fluctuations in
  the underlying nanoparticle structures, which are indicated in the
  legend, and a randomized time series.}
\label{fig-SVs}
\end{figure}

\begin{figure}[htb]
\begin{tabular}{cc} 
\resizebox{18pc}{!}{\includegraphics{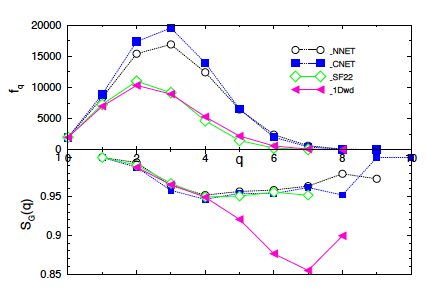}}\
\end{tabular}
\caption{The entropy $S_Q(q)$ and the number of simplexes and shared faces $f_q$
  at the  topological level $q$ of the QTS graphs of the charge
  fluctuations in the indicated  nanoparticle assemblies.}
\label{fig-entropy}
\end{figure}

\begin{widetext}
\begin{figure*}[htb]
\begin{tabular}{cc} 
\resizebox{28pc}{!}{\includegraphics{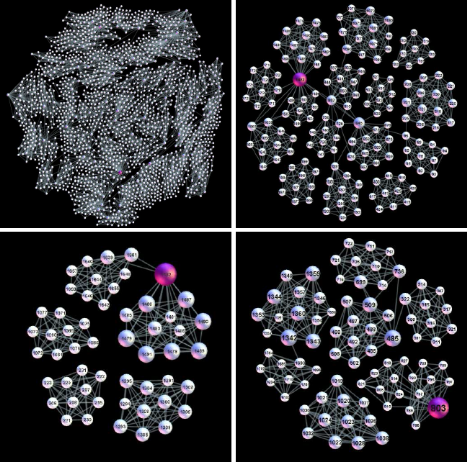}}\\
\end{tabular}
\caption{The structure of connections of the phase space manifolds corresponding to the charge fluctuations in the nanoparticle network NNET: complete QTS$_-$NNET graph (top left) and its highest topological layers  $q=8$, top right, and $q=9$, bottom left. For comparison, the layer $q=8$ of the QTS$_-$CNET graph is displayed, bottom right. In figures of the higher topological layers, ID of each node refers to the index of the time interval $t_k$ of
the time series. All edges between these nodes that exist in the
adjacency matrix are shown.}
\label{fig-QTSnets}
\end{figure*}
\end{widetext}

The time series of charge fluctuations $N_Q(t_k)$ are converted into QTS-graphs, as described in Section \ref{sec-methods}. With analysis of these graphs, we provide a robust topological description of the collective charge fluctuations in different nanoparticle assemblies. For a better comparison, we map an equal segment of each time series in the considered nanoparticle assemblies. Hence, we generate four QTS-graphs of 2000 nodes; each QTS-graph clearly relates with the underlying nanoparticle structure in Fig.\ \ref{fig-nanonets}, from which the time series is taken. The utilized mapping rules account for the strength of fluctuations; consequently, the QTS-graphs of different structure are obtained.  Fig.\ \ref{fig-AMs} displays the adjacency matrices of all four QTS-graphs. 
The pronounced block-diagonal structure in the adjacency matrices for
QTS-graphs of CNET and 1Dwd indicates the occurrence of communities of dense links. While in the case of QTS-graphs of NNET and SF22, a sparse structure of connections among the diagonal blocks appears,
which is compatible with a hierarchical organization of communities
identifiable at the graph level.
The ranking distributions of the node's degree and topological dimension of all QTS-graphs are shown in Fig.\
\ref{fig-ranked-nodes}. Separate fits for small and large rank
according to the discrete  generalized beta function are provided,
except for the case of FS22, where a satisfactory fit is obtained by a
power-law decay with a cut-off (see legends).
\begin{table}[!htb]
\caption{Standard graph theory measures ($d$--diameter,  $\langle
  k\rangle $--average degree, $\langle \ell \rangle$--average path length, Cc--clustering coefficient, the number of triangles, modularity) of the QTS-graphs representing collective charge fluctuations in nanonetworks from Fig.\ \ref{fig-nanonets}. }
\label{tab-standard-str}
\begin{tabular}{ccccccc}
QTS-graph&$d$& $\langle k\rangle $& $\langle \ell \rangle$& Cc & No.triang& modul\cr
\hline
$_-$1Dwd & 10 & 6.98 & 5.14 & 0.77& 10314 & 0.907\cr
$_-$SF22 & 10 & 7.27 & 4.89 & 0.76 & 10987 & 0.897 \cr
$_-$CNET & 9 & 8.91 & 4.66 & 0.81 & 17369 & 0.906 \cr
$_-$NNET & 11 & 8.35 & 4.86 & 0.79& 15418 & 0.909 \cr
\hline
\end{tabular}
\end{table}
A collection of standard graph-theoretic measures of all 
QTS-graphs is given in Table\ \ref{tab-standard-str}, exhibiting small
differences at the graph level.  However, the higher-order structures
of these graphs, which are described by  the algebraic topology measures, can be considerably different; the
results are displayed in  Figs.\ \ref{fig-SVs}-\ref{fig-entropy}  and
illustrated by Fig.\ \ref{fig-QTSnets}.

The higher-order structures of the QTS-graphs are revealed by
determining the components of three structure vectors, which are defined in section\ \ref{sec-methods}. The results are displayed in Fig.\ \ref{fig-SVs}.  Their numerical values are summarized in Table\ \ref{tab-vectors}.
For a comparison, we also show the results for a
graph, which is obtained from a  randomized time series; for this
purpose, such series is obtained by interchanging randomly-selected pairs of data points in the time series from NNET. 
Notably, the randomized time series results in a graph exhibiting much
simpler structure than the other graphs, which represent the fractal time series. The topological complexity of these QTS-graphs manifests in the occurrence of a vast number of topological levels $q=0,1,2\cdots
q_{max}$ in accord with the number of combinatorial spaces and their
interconnections at higher topological levels.  As an example, in Fig.\
\ref{fig-QTSnets} we display the QTS-graph of NNET assembly. In two
separate plots we also demonstrate the complexity of its top
topological levels, $q=8$ and $q=9$. 
Specifically, it contains eight  10-cliques; three 10-cliques are
separated, and the other five are
interconnected via 9-cliques at the lower level $q=8$. Here, according
to the table\ \ref{tab-vectors},  36-8=28 additional  9-cliques exist
and are interconnected such that  33 components occur at the level
$q=8$. That is, 33-28=5 components are identifiable at the upper
level, cf. Fig.\ \ref{fig-QTSnets} bottom left.
Considering the level below, $q=7$, gives 105-36=69  new
8-cliques and 
86 components, which suggests that maximally 86-69=17
groups of nodes can be distinguished at the level $q=8$.  
Among these, five groups contain higher cliques, leaving at most 12
groups that are visually distinct  in the top right figure \ref{fig-QTSnets}.
In contrast, the single 11-clique in QTS-graph of CNET remains isolated down to the level $q=7$ (cf. table\
 \ref{tab-vectors}). Together with  17  other 9-cliques that appear at the level $q=8$, they make 18 components. At the
 level below, we identify  137-18 =119 new  8-cliques. Then comparing
 the number of components, 128-119=9 suggests that the cliques at the
 upper level $q=8$ make at most nine groups, one of which refer to the
 11-clique of the top levels. This structure is also shown in Fig.\
 \ref{fig-QTSnets} bottom right.

According to Fig.\ \ref{fig-SVs}, the second structure
vectors of QTS-graphs of SF22 and 1Dwd exhibit certain similarity.
Interestingly enough, the presence of charge disorder in the linear chain
of nanoparticles leads to a more complex QTS-graph than the SF22
assembly. Namely, the charge disorder blocks the only existing path in
the linear chain, which results in a sequence of zero entries in the
time series, which is mapped onto a large clique linked to two or more
near non-zero entries. In contrast, blocking a linear segment in SF22 structure causes tunnelings along the alternative paths. Hence, no abrupt changes in the time series occur. Apart from the high topology
levels, the similarity between the QTS-graphs of NNET and CNET, on one side, and SF22 and 1Dwd, on the other, is apparent at
lower $q$-levels, cf. Fig.\ \ref{fig-SVs}. These findings also apply
to the topological response function $f_q$ and the entropy $S_Q(q)$,
shown in Fig.\ \ref{fig-entropy}. 
The occurrence of mutually isolated cliques, i.e, at $q=1$ or $q=q_{max}$, leads to the
occupation probability $p_q^i\lesssim 1$ and the entropy close to
zero. The simplicial complexes that occur at the intermediate $q$-levels share the faces at the level $q-1$. Hence, same nodes participate in the identified complexes, leading to a lower occupation probability of the level and  the entropy drops, as shown in
Fig.\ \ref{fig-entropy}. The 2-dimensional nanoparticle assemblies exhibit complex QTS-graphs in which a large number of simplicial complexes, sharing many nodes,  is observed. In contrast, a small number of such compounds occur in the case of the linear chain structure with charge disorder. Thus, a reduced number of nodes are involved; this situation manifests in a more pronounced entropy minimum at $q_{max}-1$ than in the other graphs.

\section{Conclusions\label{sec-conclusions}}
Using the extensive numerical study at different scales, we have
demonstrated how the structure of tunneling junctions in the arrays
of metallic nanoparticles affects the  collective nature of the charge
transport within the Coulomb-blockade regime.
The simulations of single-electron tunnelings in nanoparticle
arrays of different architecture are combined with the fractal analysis of time series and algebraic topology techniques applied to the abstract graphs in the state space.
We quantitatively describe the aggregate fluctuations at the global
scale; the collective dynamics builds on communication among local
events, which is enabled by the structural features of the nanoparticle
network with junctions.  Two structural components can locally affect the formation of the conducting paths between the electrodes encasing the array.  These are the branching at each nanoparticle, which increases geometrical entanglement of the conduction paths,  and the presence of linear segments whose tunneling conduction is better understood,  which reduces it. 

The comparative analysis of different nanonetworks in conjunction with the size of electrodes leads to the following general conclusions.
\begin{itemize}
\item A higher level of cooperation in the charge fluctuations through the sample that is measured by the fractal indicators of the time series is consistent with a larger topological complexity of the phase space manifolds yielding an enhanced $I(V)$ nonlinearity beyond the voltage threshold.
\item A suitable combination of the local branching and global topological defects that allow a formation of large draining basins, as illustrated by NNET structure in our analysis, enables an enhanced collective dynamics.

\item The size of electrodes can considerably influence the strength
  of aggregate fluctuations, and thus the structure of the complex
  phase space and the $I(V)$ curve. In particular, point-size
  electrodes enhance the formation of several nearly-shortest paths;
  they can be physically close to each other in arrays of a compact
  structure with an extensive branching along the main path.  A good
  example is CNET studied in this work. On the other hand, the
  presence of significant linear segments, such as SF22, increases the
  difference between the path lengths, which reduces the probability
  of a coordinated drainage. In such cases, extended electrodes allow
  the formation of several channels of a similar structure.
\end{itemize}
Our analysis also sheds a new light onto the effects of the charge
disorder on the system's conduction capabilities. Both the temporal
correlations and the topology of phase space manifolds demonstrate
dramatic effects of charge disorder in the linear chain of nanoparticles, which comprises a single conduction path.  In a two-dimensional array, such as SF22, blocking some linear segments of the structure by charge disorder affects the morphology of conduction basins, resulting in a reduced current. 

The presented results reveal how the architecture implicates  the emergent conduction features of nanoparticle assemblies.  They may serve as a guide for the design of the nanoparticle devices with improved conduction.
We have not explicitly considered the effects of finite temperature
and varied widths of the tunneling junctions, which often appear in
the experiments. They also may alter the tunneling process at the local scale and,  according to this work, imply changes in the  collective charge fluctuations.  
Apart from these practical aspects, our work stresses the importance
of different modeling approaches. In this context, the methods of
algebraic topology applied to the graphs in the system's phase space
are opening a new perspective for the analysis of fluctuations at a nanoscale.

\section*{\normalsize Acknowledgments}
This work was supported by the Program P1-0044 of the research agency
of the Republic of Slovenia and in part by the Projects OI 174014  and
OI 171037  and III 41011 by
the Ministry of Education, Science and Technological Development of the Republic of Serbia. MA
also thanks for kind hospitality during his stay at the
Department of Theoretical Physics, Jo\v zef stefan Institute, where
this work was done.

\appendix
\begin{table*}[!tbp]
\caption{Components of three structure vectors of the graphs
  representing  charge-fluctuation time series in nanonetworks from
  Fig.\ \ref{fig-nanonets}.}
\label{tab-vectors}
\begin{tabular}{|c|ccc|ccc|ccc|ccc|ccc|}
\hline
  QTS$_-$ & \multicolumn{3}{c|}{$_-$NNET}                   &
                                                   \multicolumn{3}{c|}{$_-$CNET}
  & \multicolumn{3}{c|}{$_-$SF22}                   &
                                                  \multicolumn{3}{c|}{$_-$1Dwd}
  & \multicolumn{3}{c|}{$_-$randomised TS}                   \\ \hline
q  & $G_q$ & $n_q$ & \multicolumn{1}{c|}{$\hat G_q$} & $G_q$ & $n_q$ & \multicolumn{1}{c|}{$\hat G_q$} & $G_q$ & $n_q$ & \multicolumn{1}{c|}{$\hat G_q$} & $G_q$ &$n_q$ & \multicolumn{1}{c|}{$\hat G_q$} & $G_q$ & $n_q$ & \multicolumn{1}{c|}{$\hat G_q$} \\ \hline
0  & 1   & 1458  & 0.99                          & 1   & 1550  & 0.99                          & 1   & 1684  & 0.99                          & 1   & 1721  & 0.99                          & 1   & 1403  & 0.99                          \\ \cline{1-1}
1  & 10  & 1458  & 0.99                          & 6   & 1550  & 0.99                          & 7   & 1684  & 0.99                          & 7   & 1721  & 0.99                          & 13  & 1403  & 0.99                          \\ \cline{1-1}
2  & 265 & 1457  & 0.82                          & 193 & 1549  & 0.88                          & 455 & 1683  & 0.73                          & 503 & 1721  & 0.71                          & 904 & 1400  & 0.35                          \\ \cline{1-1}
3  & 497 & 1326  & 0.63                          & 399 & 1473  & 0.73                          & 673 & 1410  & 0.52                          & 733 & 1448  & 0.49                          & 766 & 873   & 0.12                          \\ \cline{1-1}
4  & 570 & 1025  & 0.44                          & 594 & 1238  & 0..52                         & 638 & 935   & 0.31                          & 517 & 893   & 0.42                          & 265 & 283   & 0.06                          \\ \cline{1-1}
5  & 430 & 628   & 0.32                          & 512 & 826   & 0.38                          & 346 & 418   & 0.17                          & 268 & 433   & 0.38                          & 67  & 69    & 0.03                          \\ \cline{1-1}
6  & 252 & 313   & 0.20                          & 351 & 438   & 0.20                          & 103 & 119   & 0.13                          & 145 & 179   & 0.19                          & 12  & 12    & 0                             \\ \cline{1-1}
7  & 86  & 105   & 0.18                          & 128 & 137   & 0.07                          & 25  & 25    & 0                             & 29  & 37    & 0.22                          & 1   & 1     & 0                             \\ \cline{1-1}
8  & 33  & 36    & 0.08                          & 18  & 18    & 0                             &     &       &                               & 9   & 9     & 0                             &     &       &                               \\ \cline{1-1}
9  & 8   & 8     & 0                             & 1   & 1     & 0                             &     &       &                               &     &     &                             &     &       &                               \\ \cline{1-1}
10 &     &       &                               & 1   & 1     & 0
         &     &       &                               &     &      &
                                                                                                     &     &       &                               \\ \cline{1-1}
\hline
\end{tabular}
\end{table*}

\end{document}